\documentclass{emulateapj}
\usepackage{apjfonts}
\usepackage{epsfig}

\submitted{ApJ, accepted}

\shorttitle{HST-1 in the M87 Jet}
\shortauthors{Harris et al.}

\begin{document}
\title{The Outburst of HST-1 in the M87 Jet}

\author{D. E. Harris}
\affil{Smithsonian Astrophysical Observatory, 60 Garden St.,
Cambridge, MA 02138} 
\email{harris@cfa.harvard.edu}

\author{C. C. Cheung\altaffilmark{1}}
\affil{MIT Kavli
Institute for Astrophysics \& Space Research, 77 Massachusetts Ave.,
Cambridge, MA 02139}

\author{J. A. Biretta and W. B. Sparks}
\affil{Space Telescope Science Institute, 3700 San Martin Drive,
Baltimore, MD 21218}

\author{W. Junor}
\affil{Space \& Atmospheric Sciences, Los Alamos National Laboratory,
P.O. Box 1663, Los Alamos, NM 87545}

\author{E. S. Perlman}
\affil{Joint Ctr. for Astrophysics, University of Maryland, Baltimore
County, 1000 Hilltop Circle, Baltimore, MD 21250}

\author{A. S. Wilson}
\affil{Astronomy Department, University of Maryland, College Park, MD 20742}

\altaffiltext{1}{Jansky Postdoctoral Fellow; National Radio Astronomy
Observatory.  Now hosted by Kavli Institute for Particle Astrophysics and
Cosmology, Stanford University, Stanford, CA 94305}


\begin{abstract}

The X-ray intensity of knot HST-1, 0.85$^{\prime\prime}$ from the
nucleus of the radio galaxy M87, has increased by more than a factor
of 50 during the last 5 years.  The optical increase is similar and
our more limited radio data indicate a commensurate activity.  We give
the primary results of our Chandra X-ray Observatory monitoring
program and consider some of the implications of this extreme
variability in a relativistic jet.  We find that the data support a
'modest beaming synchrotron' model as indicated in our earlier
papers.  Based on this model, the decay of the X-ray lightcurve 
appears to be dominated by light travel time across the emitting
region of HST-1, rather than synchrotron loss timescales.

\end{abstract}



\section{Introduction}  

This paper is the third from a project to monitor the M87 jet with the
Chandra X-ray Observatory with complimentary data from the Hubble
Space Telescope (HST), and more recently with the Very Large Array
(VLA) and the Very Long Baseline Array (VLBA)\footnote{The National
Radio Astronomy Observatory is a facility of the National Science
Foundation operated under cooperative agreement by Associated
Universities, Inc.}.  The main goal of the project was based on the
premise that synchrotron X-ray emission would come from electrons with
halflives of order a year or less whereas inverse Compton X-ray
emission would come from very low energy electrons with halflives
exceeding $10^4$ years \citep{hk02}.  In \citet{pap1} (hereinafter
'Paper I') we reported the results from the first year, 2002Jan-Jul.
The main findings were significant intensity variability of the core
(expected), and of HST-1, a knot in the jet 0.85$''$ (65pc projected)
from the core.  We argued that the intensity changes were larger for
the harder X-ray bands and suggested a synchrotron loss model with
modest beaming (Doppler factor, $\delta$, of order 3 or 4, in agreement
with earlier proper motion studies; \citet{bir99}).  In \citet{pap2} (Paper
II), it was suggested that the timescale for optical decay of the
lightcurve for HST-1 was similar to the X-ray decay time, and thus not
the factor of 10 longer which was predicted by our original
synchrotron loss model.  Since the measured decay time was based on
only a single minor change of optical intensity for each color, we
consider this question still open.

In this paper we concentrate on the recent Chandra data from 2003-2005
but will also include optical and radio data for HST-1 as required,
leaving the details to other papers in preparation.  The 'giant flare'
of HST-1 (the X-ray intensity increased by a factor of 50 in 4 years)
has forced us to abandon our original methods of data reduction which
were devised the first year of our monitoring when ACIS pileup was not
a problem.  [`ACIS' is the acronym for the Advanced CCD Imaging
Spectrometer and `pileup' is discussed in section~\ref{sec-pileup}.]
After a short section on the observations (\S~2) we deal
with the data reduction methods and give the basic results for each
type of measurement (\S~3).  We review the synchrotron
parameters in \S~4, and discuss implications
of the lightcurve timescales in \S~5.

We take the distance to M87 to be 16 Mpc \citep{ton91} so that one
arcsec~=~77pc.  We use the conventional definition of spectral index,
$\alpha$: flux density, S$_{\nu}~\propto~\nu^{-\alpha}$.

\begin{figure*}
\begin{center}
\epsfig{file=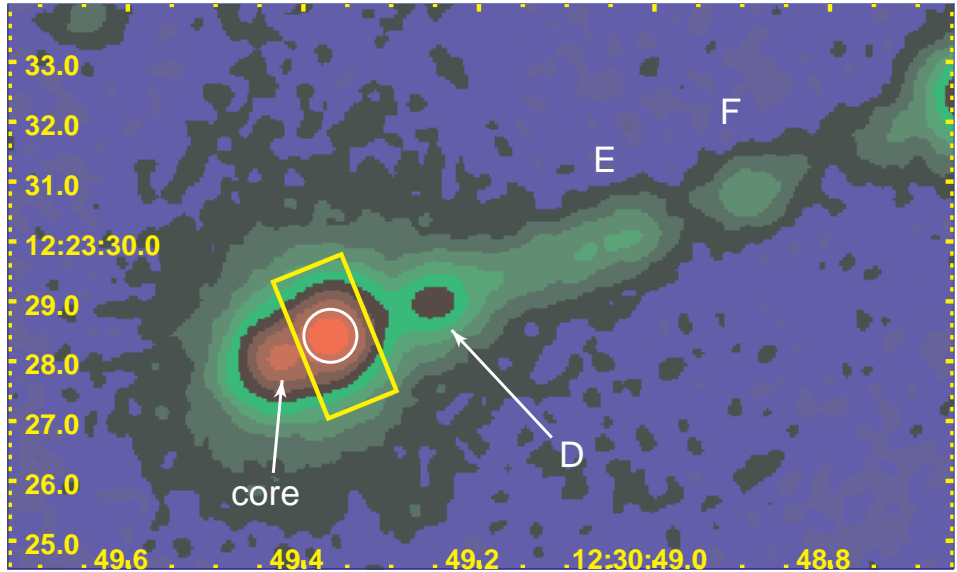,width=5in,angle=0}
\end{center}
\caption{The inner section of the M87 X-ray Jet: a stacked image from
  epochs A to V with an effective exposure time of order 110ks.  The
  data come from evt2 files so the true (average) intensity of HST-1   
  is underestimated because of pileup.  The coordinates labels are   
  Right Ascension and Declination in J2000.  The intensity units are  
  electron volts per second per 0.0492 arcsec pixel, and the peak   
  value is 5.5.  The faintest red corresponds to 0.5 and the black
  below the green is close to 0.01.  The yellow rectangle around HST-1
  is that used for the lightcurve photometry and the small circle is
  that used in PapI (sections 3.2, 3.3.).
  \label{fig:regions}}
\end{figure*}

\section{Observations\label{sec-obs}}  

The first Chandra observations of M87 that specifically targeted the
jet were those done by Wilson in 2000 Jul \citep{wy02,per05}.  They
experimented with various ACIS frame times in order to minimize the
effects of pileup (two or more photons arriving at the same pixel
during the same frame time, thus masquerading as a single event with
the sum of the energies; see, e.g. \citet[][]{dav01}).  They found that
a frame time of 0.4s gave essentially the same countrate as 0.1s for
the nucleus of M87 which, at that time, was the brightest feature.
Thus we adopted the same ACIS setup: chip S3 (back illuminated for low
energy response), 1/8th subarray, and 0.4s frame time.  A list of our
observations is given in Table~\ref{tab:obs}. Basically, we have been
getting a 5 ks observation approximately every 6 weeks for four
'seasons': mid November to early August (when M87 is far enough from
the Sun to be observable with Chandra and the HST).  Our approved
proposal for Chandra AO7 will continue this strategy (2005Nov to
2006Aug).  During the 2005 season we have obtained an extra 5
observations designed to give us weekly coverage for 2 months to check
for shorter timescale changes.

\begin{table}
\begin{center}
\caption{Chandra dates and X-ray intensities of HST-1\label{tab:obs}}
\begin{tabular}{llccccr}
\tableline\tableline
\multicolumn{4}{c}{Observational Parameters} & \multicolumn{3}{c}{Intensities of HST-1} \\
Epoch 	&Date 
& Obsid	&Livetime 
& K$^{\rm a}$ 	&f(0.2-6keV)$^{\rm b}$  
& S(2keV)$^{\rm c}$\\
label 	&  
&  		&(sec) 
& (keV/s) 	&(10$^{-12}$ cgs)
& (nJy)\\
\tableline
A$^{\rm d}$ & 2000Jul30  & 1808  & 12845 & 0.247 & 0.729 & 29.2	\\
B   & 2002Jan16  & 3085  & 4889   &  0.734   &  2.32    &   92.8  \\
C   & 2002Feb12  & 3084  & 4655   &  0.588   &  1.87    &   74.8  \\
D   & 2002Mar30  & 3086  & 5089   &  0.576   &  1.84    &   73.6  \\
E   & 2002Jun08  & 3087  & 4973   &  0.798   &  2.57    &  102.8  \\
F   & 2002Jul24  & 3088  & 4708   &  1.096   &  3.54    &  141.6  \\
G   & 2002Nov17  & 3975  & 5287   &  0.799   &  2.62    &  104.8  \\
H   & 2002Dec29  & 3976  & 4792   &  0.653   &  2.14    &   85.6  \\
I   & 2003Feb04  & 3977  & 5276   &  0.645   &  2.12    &   84.8  \\
J   & 2003Mar09  & 3978  & 4852   &  0.872   &  2.88    &  115.2  \\
K   & 2003Apr14  & 3979  & 4492   &  1.071   &  3.54    &  141.6  \\
L   & 2003May18  & 3980  & 4788   &  1.022   &  3.39    &  135.6  \\
M   & 2003Jul03  & 3981  & 4677   &  0.859   &  2.86    &  114.4  \\
N   & 2003Aug08  & 3982  & 4841   &  1.214   &  4.06    &  162.4  \\
O   & 2003Nov11  & 4917  & 5028   &  2.192   &  7.36    &  294.4  \\
P   & 2003Dec29  & 4918  & 4677   &  2.041   &  6.88    &  275.2  \\
Q   & 2004Feb12  & 4919  & 4703   &  4.079   &  13.81   &  552.4  \\
R$^{\rm e}$ & 2004Mar29 & 4920 & 5235 & .. & .. & ..  \\
S   & 2004May13  & 4921  & 5251   &  5.358   &  18.23   &  729.2  \\
T   & 2004Jun23  & 4922  & 4543   &  5.323   &  18.16   &  726.4  \\
U   & 2004Aug05  & 4923  & 4633   &  5.636   &  19.28   &  771.2  \\
V   & 2004Nov26  & 5737  & 4237   &  7.494   &  25.76   & 1030	   \\
W   & 2005Jan24  & 5738  & 4666   &  8.316   &  28.66   & 1146	   \\
X   & 2005Feb14  & 5739  & 5154   &  8.785   &  30.30   & 1212	 \\
Ya  & 2005Apr22  & 5740  & 4699   &  12.42   &  42.91   & 1716	   \\
Yb  & 2005Apr28  & 5744  & 4699   &  12.17   &  42.08   & 1683	   \\
Yc  & 2005May04  & 5745  & 4705   &  11.80   & 	40.81   & 1632	  \\
Yd  & 2005May13  & 5746  & 5142   &  11.60   & 	40.14   & 1606	  \\
Ye  & 2005May22  & 5747  & 4701   &  11.24   &  38.92   & 1557	   \\
Yf  & 2005May30  & 5748  & 4699   &  10.66   &  36.92   & 1477	   \\
Yg  & 2005Jun03  & 5741  & 4698   &  10.43   &  36.14   & 1446     \\
Yh  & 2005Jun21  & 5742  & 4703   &  10.27   &  35.58   & 1423  \\
Yi  & 2005Aug06  & 5743  & 4672   &  7.105   &  24.64   & 985.5 \\
\tableline
\end{tabular}
\end{center}
Note. -- All observations had 1/8th subarray ACIS-S7 chip only,
and 0.4s frame time.\\
$^{\rm a}$ The values of K are our primary measured quantity.
They come from summing the energies of all events (from the 'evt1'
file) within a rectangle of length 5 pixels transverse to the jet
and 2.5 pixels along the jet (fig.~\ref{fig:regions}).  The upstream
edge is approximately midway between HST-1 and the core component.
No background subtraction was employed: when the source is weak, the
background (in counts) is of order 1\%, and for the highest
intensity, the background is less than 0.3\%.  There is no
correction for the buildup of ACIS contamination.\\
$^{\rm b}$ f(0.2-6keV) is the observed X-ray flux (the flux
density integrated from 0.2 to 6 keV) with units of 10$^{-12}$
erg~cm$^{-2}$~s$^{-1}$.  An approximate correction has been made for
ACIS contamination buildup but not for Galactic absorption.\\
$^{\rm c}$ The flux density at 2 keV has been calculated from
the observed flux assuming a spectral index, $\alpha$=1.5.  For
$\alpha$=1.75, the listed values should be reduced by a factor 0.75
and for $\alpha$=1.25, they should be increased by a factor of
1.25. 1 nJy is 10$^{-32}$ erg~cm$^{-2}$~s$^{-1}$~Hz$^{-1}$.\\
$^{\rm d}$ Archival data from \citet{wy02}.\\
$^{\rm e}$ This observation was taken in continuous clocking
mode, so there is no 2D image available.
\end{table}

\section{Data reduction and basic results\label{sec:data}} 

While we have maintained a fixed observing strategy, the data
reduction has been done over the course of several years and thus
the software reduction package, 
`CIAO'\footnote{\url{http://asc.harvard.edu/ciao/}} (Chandra
Interactive Analysis of Observations), has evolved through
several releases, as has the calibration database.  However, all data
except the 'keV/s' values have been corrected for the contamination on
the ACIS detector, and we are not aware of any software or calibration
changes that would adversely affect our results.

The essential preparatory procedures are removal of the pixel
randomization and registration to align the X-ray core with the radio
core.  After setting up the bad pixel file and checking the lightcurve
of a large background region for flares (since we use small extraction
regions and no large flares were found, we did not time filter any
observation), we removed the pixel randomization imposed by the
standard pipeline by following the relevant CIAO thread.  Registration
of the X-ray image was accomplished by measuring the apparent
positions of the core and knot A on both radio and X-ray images (the
latter binned to 1/10th native ACIS pixel and smoothed with a Gaussian
of FWHM=0.25$''$), and then changing the values of the FITS header
keywords in the Chandra event file.  In general, this procedure
introduces an uncertainty of approximately 0.05$''$ or less for the
early part of the program when the intensity of the core was
comparable to that of HST-1.  However, when HST-1 became more than 10
times brighter than the core, our ability to determine the centroid of
the core emission is compromised, and we then gave more weight to the
position of knot A which is not so well determined because knot A is
resolved.

\subsection{Intensities from fluxmaps}
The original data reduction procedure is outlined in Paper I and
involves constructing monochromatic exposure maps for three energy
bands. Dividing the data by the exposure map we obtain 3 flux maps.
Each flux map was then multiplied by $h\nu_{\rm nom}$ to obtain units of
ergs cm$^{-2}$ s$^{-1}$.  $\nu_{\rm nom}$ corresponds to the nominal
energy for each band (chosen to approximate the band center weighted
by the effective area) which was used to construct the relevant
exposure map. This permits us to measure fluxes with normal photometric
procedures.  All that is required is to sum the flux within a
measuring aperture and multiply the result by the ratio of the mean
energy of events within the aperture to the nominal energy of the band
being measured.  We have redone the fluxmaps now that CIAO
automatically corrects the exposure maps for ACIS contamination
buildup.  Previously we had used a fixed region of M87's thermal hot
gas emission to make this correction.

When using small apertures (our standard region was a circle with
r=0.44$''$ in order to separate the core from HST-1; see
fig.~\ref{fig:regions}) a point spread function (PSF) correction of
1.55$\pm$0.05 is required to 'recover' the fraction of the PSF outside
the small circle.  Although it is known that the PSF size is a
function of energy, from our PSF simulation with ChaRT/Marx, we found
values between 1.50 and 1.60 for our 3 energy bands.  From the same
simulations, we find that 6$\pm$4\% of each PSF falls within the
adjacent measuring aperture (this is for the special case of the small
circles for the core and HST-1).  In Paper I, these corrections had
not yet been devised, so the flux scales in the figures should not be
used to derive absolute values.  We provide our best flux estimates
later in this paper.

\subsection{Intensities from piled up data: keV/s}
\label{sec-pileup}

Pileup for CCD's occurs when two or more events are received for the
same pixel during a single frame time \citep{dav01}.  Thus we lose the inherent
spectral energy distribution one normally associates with ACIS data.
However, to first order, the energy content of the arriving photons is
preserved and a measure of the total band is still available.  To
obtain intensities for lightcurves, we first followed the 'afterglow
thread'\footnote{\url{http://cxc.harvard.edu/ciao/threads/acisdetectafterglow/}.
Note however that with the advent of pipeline version DS7.4, this
procedure is no longer required.}  to recover the events with grade
migration that caused them to be mistaken for afterglow events.  Then
we checked to see that all events of interest would be accepted by a
suitable energy filter.  The incident spectrum of HST-1, both as
measured when the knot was weak (essentially no pileup) and from our
observation in continuous clocking mode, is essentially contained in
the band 0.2 to 6 keV (probably 95\% of the counts are below 4 keV).
To recover the energy associated with piled events that now appear
above 6 keV, we examined the event file at the higher energies and set
the upper cutoff to a value that includes all events associated with
HST-1 (up to 17 keV in some cases).

\begin{figure*}
\begin{center}
\epsfig{file=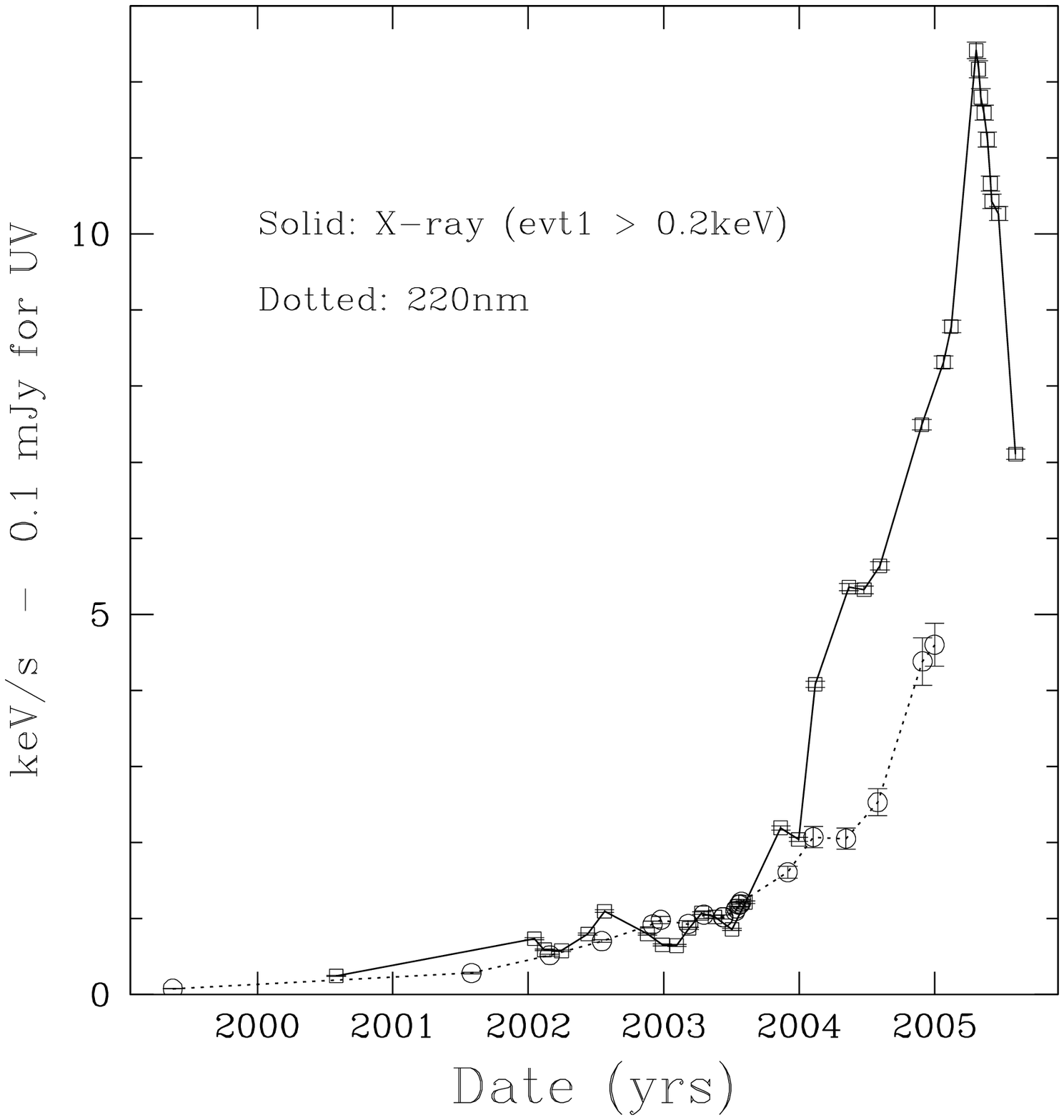,width=3.0in,angle=0}
\epsfig{file=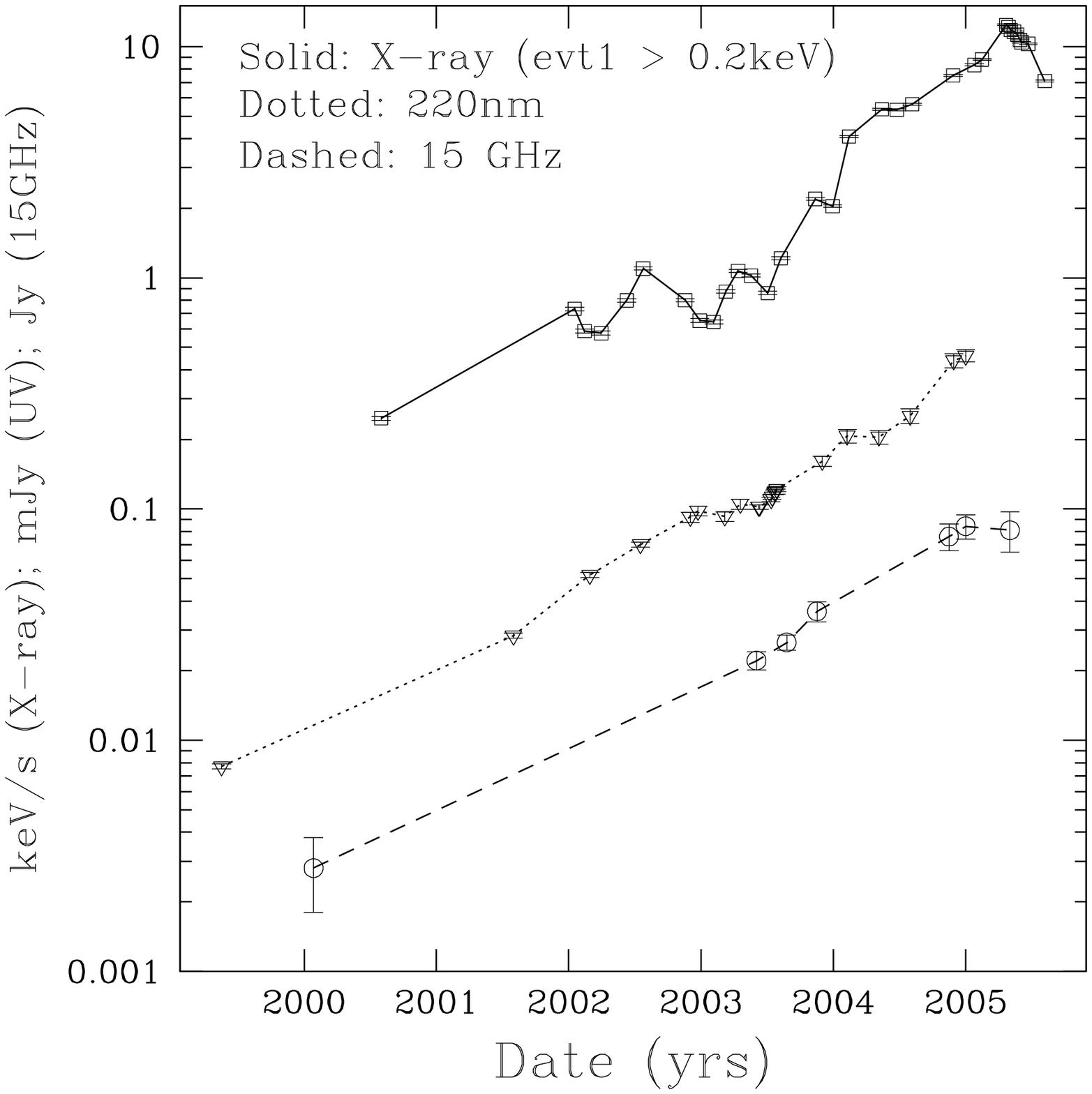,width=3.0in,angle=0}
\end{center}
\caption{Multi-wavelength lightcurves for knot HST-1 in the M87 jet.
The left panel shows the X-ray, and UV lightcurves on a linear scale
to more clearly show the higher levels.  On the right is the same plot
on a log scale in order to better compare slopes (percentage changes).
For this version we have added radio data at 15 GHz (the 2000 datum
comes from an archival VLA observation).  Both X-ray curves are from
the evt1 files in keV/s.  For the UV data, the left panel has units of
0.1 mJy whereas on the right, the units are mJy.
  \label{fig:hst1lc}}
\end{figure*}

The CIAO tool 'dmstat' then gives the total energy (in eV) of all
events.  Dividing this by 1000$\times$livetime (where `livetime' is
the total time the detector is accumulating data) provides an
intensity measurement in keV/s.  Through early 2004, pileup was fairly
mild, so we continued measuring the event2 file. As the pileup
increased, we found we were losing more and more events to grade
migration so the data shown in fig.~\ref{fig:hst1lc} are based on the
event 1 file with no grade filtering and only the standard good time
intervals applied.  A rectangular extraction region (see
fig.~\ref{fig:regions}) was used for this photometry.

There are several limitations to this method.  It is impossible to get
spectral information like band fluxes; there is no {\it a priori} method to
correct for ACIS contamination
buildup\footnote{\url{http://cxc.harvard.edu/ciao/threads/aciscontam/}};
and there may have been bona fide events that were rejected by on-line
software aboard the satellite and thus are unrecoverable.  In spite of
these deficiencies, this 'keV/s' method provides a reasonably direct
estimate of the total observed intensity.

\subsection{Problems Associated with the Spectral Analyses\label{sec:spectral}}

For spectral analyses of HST-1 there are between several hundred and
more than 3000 counts in a typical 5ks observation.  We have performed
spectral fits for simple power laws (PL), power laws with a high
energy cutoff (cutPL), and both models in conjunction with the
'jdpileup model' \citep{dav01}, as implemented in {\it
Sherpa} \citep{fre01}.  There are a number of problems.

The jdpileup model, even with the 'Monte-Powell' method which is
advertised as being the best option to locate the true minimum in
chi-squared, can produce a best fit which is obviously wrong:  the fit
gives an amplitude which is up to a factor of 10 too large, with an
excessively large pileup fraction.  This formal fit occurs because the
fit finds a solution on the heavily piled up section of the pileup
curve \citep[fig.~6]{dav01}.  This is particularly evident
for featureless spectra such as power laws.  In the fitting algorithm,
there is a 'degeneracy' for featureless spectra between the power
law exponent and the pileup 'survival probability' parameter, jdp.alpha.

The only way we found to mitigate this problem was to constrain
several fit parameters: Nh, the total column density of neutral
hydrogen per square cm towards the source, was forced to be at least
the Galactic value of 2.5$\times10^{20}$~cm$^{-2}$ \citep{sta92}; the
survival probability parameter jdp.alpha.min=0.2; and the spectral
index, $\alpha$, was constrained to lie between 0 and 3.  It was also
necessary to pick good values for the initial guess for jdp.alpha,
$\alpha$, and the amplitude of the power law.

Another uncertainty is the mutual dependence between Nh and $\alpha$
which can be demonstrated by a simple example using the spectral file
for HST-1 from epoch D.  The power law fit gives Nh=(3.4, +1.9,
-1.5)$\times10^{20}$cm$^{-2}$ and $\alpha$=1.46$\pm$0.14.  Freezing Nh
to values of 2.5, 3.5, and 4.5$\times$10$^{20}$, results in $\alpha$ values of
1.39$\pm$0.07, 1.47$\pm$0.07, and 1.54$\pm$0.08 respectively.
Although it is conceivable that Nh could vary because HST-1 is at a
projected distance of only 65pc from the nucleus of M87 and its
diameter is less than a parsec, we assume it is constant, possibly
augmented by an excess absorption above the Galactic value,
i.e. intrinsic to M87.  From spectral fits before pileup became a
problem, we find fit values of Nh which average
3.5$\times10^{20}$cm$^{-2}$ (i.e. a bit larger than the Galactic
value), although each individual solution is consistent, within the
errors, with the Galactic value.

A further effect complicates the solution.  The
pileup effect we call 'Eat Thy Neighbour' comes into play for two
close sources.  If a core photon arrives during the same frame time as
a photon from HST-1, often they will be within each other's 3x3
pixel region and thus be interpreted as a single event at the location
of the pixel with the larger energy.

Finally, the value of $\alpha$ depends on the choice between a simple
power law and a power law with an exponential cutoff.
Although in many cases the value of the cutoff energy is very large
and the $\alpha$'s of PL and cutPL are sensibly the same, for other
cases for which the fit value of the cutoff energy is below 12 keV,
the corresponding values of $\alpha$ for the two models diverge with a
concomitant shift in the fit value of Nh (an example is given in
Table~\ref{tab:cc}).


\begin{table}[b]
\begin{center}
\caption{Spectral Parameters for the CC mode data on HST-1\label{tab:cc}}
\begin{tabular}{lccc}
\tableline\tableline
Model &  Nh & $\alpha$ &$\chi^2$ \\
& 10$^{20}$cm$^{-2}$& & \\
\tableline
power law (PL)  & 7.3$\pm$0.9 & 1.72$\pm$0.06 & 0.42 \\
PL (BGsubtract) & 6.6$\pm$0.9 & 1.73$\pm$0.06 & 0.40 \\
PL + cutoff     & 2.5$\pm$0.4 & 1.18$\pm$0.11 & 0.51 \\
\tableline
\end{tabular}
\end{center}
Note. -- For the PL + cutoff, the high energy cut value is
3.7$\pm$1.1 keV and the value of Nh is the lowest allowed (set to
the Galactic value).
\end{table}

Considering these difficulties, we believe it would be premature to
present a table of spectral parameters.  For the 2002 to 2003 data
(prior to the large intensity increase), with Sherpa we find
$\alpha\approx$1.5$\pm$0.3, and Nh appears to be consistent with, or a
bit larger than, the Galactic value.  A joint fit of these data with
XSPEC constraining Nh to a common value but allowing amplitude and spectral
index to vary for each observation, results in Nh=(2.6$\pm$0.5)$\times$10$^{20}$~cm$^{-2}$ and
$\alpha$ ranging from a minimum of 1.1 to a maximum of 1.5.  It is our
opinion that with spectral fitting, we cannot confidently claim that the
spectrum of HST-1 is harder or softer when the intensity is high.

Given the sizable uncertainty in spectral parameters of HST-1 from
piled up data, we proposed a 10ks observation for Director's
Discretionary Time to obtain a continuous clocking (CC) mode
observation to be made during those few days in 2004March when the CCD
readout direction is orthogonal to the jet.  The idea was that we
could obtain one set of reliable spectral parameters and compare them
to those obtained from our normal 5 ks observation which was already
scheduled for the last week of March.  Our proposal was not granted
time, but we were allowed to change the mode of our normal 5 ks
observation to CC mode.

In CC mode pileup is not a problem since the readout is continuous and
each pixel is effectively moved every 41~$\mu$s.  We followed the
thread to correct the time of each event from the readout time to the
arrival time, thus ensuring that the correct aspect solution would be
applied.  A profile along an East-West line (i.e. the jet is projected
onto the RA axis from its normal position angle 20$^{\circ}$ away) is
shown in fig.~\ref{fig:profile}.  We then isolated knot A, HST-1, and
the core with thin rectangles on the 1D 'image' of the event file.
Notwithstanding the fact that there are no accurate calibration files
for CC mode, we generated 'pha files' (the collected events with their
energies) and corresponding effective area and redistribution matrices
to permit us to run normal Sherpa spectral analysis.  Background areas
on each end of the jet were used.  The spectral results are given in
Table~\ref{tab:cc}.

\begin{center}
\figurenum{3}
\epsscale{0.95}
\plotone{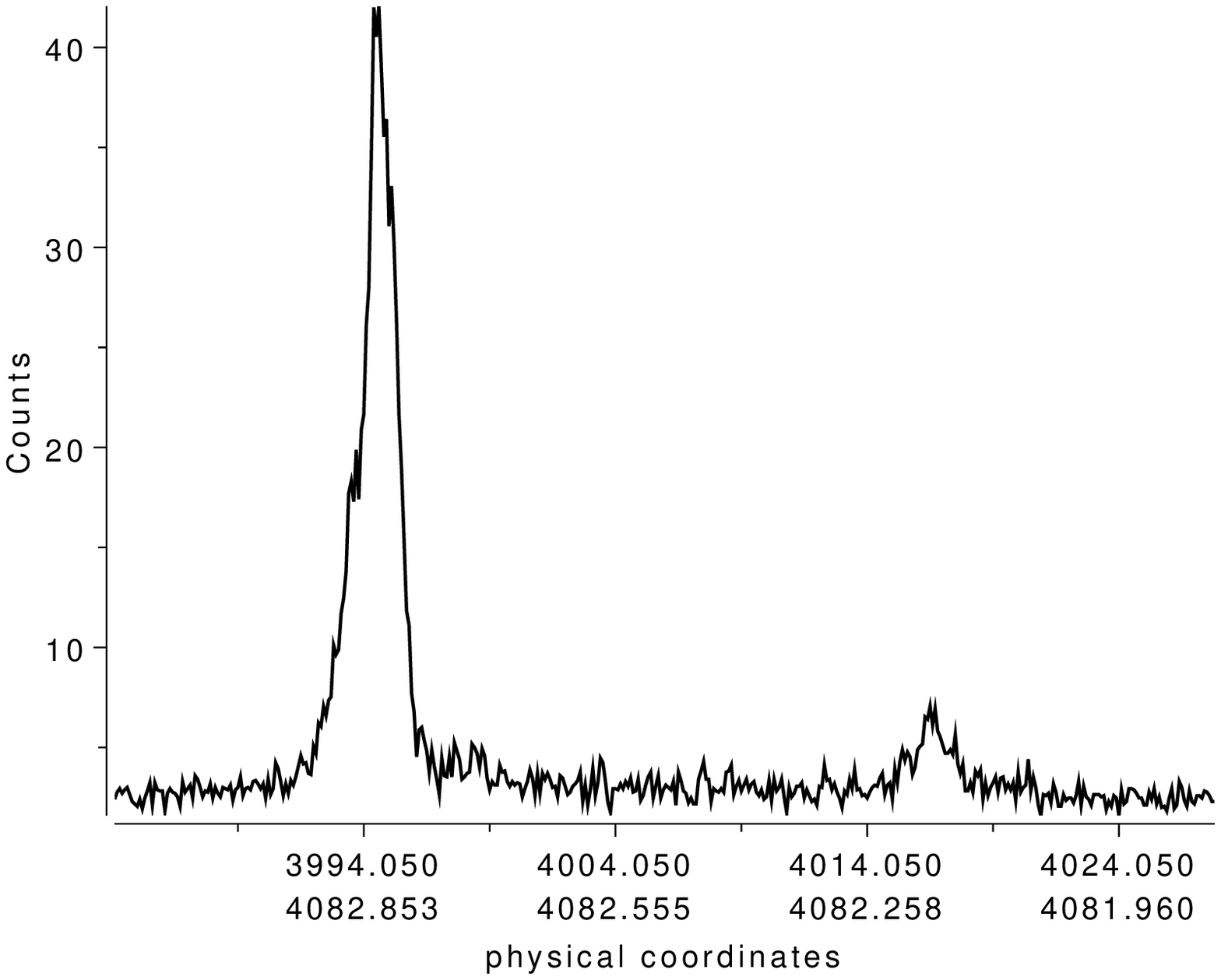}
\figcaption{The profile of the jet along the RA axis of the observation
of 2004Mar29 (Epoch R) taken in the continuous clocking mode.  The
resolution is 0.1 native ACIS pixels, or 0.0492$''$.  Emission from
the core, with an intensity similar to knot A (the feature towards
the right), is lost on the left flank of HST-1, the dominant
feature.  The location of the core should be 1.6 pixels to the left
of HST-1.  Note the relatively poor s/n of knot A, since the profile
contains the background from the whole chip.\label{fig:profile}}
\vspace{0.1in}
\end{center}

The CC mode observation was meant to confirm the jdp spectral
parameters, but insofar as $\alpha$ is concerned, the CC fit gives a
significantly larger value than the preceeding and following values.
At this time, we do not know how to decide if the CC value is wrong
because of unreliable calibration files, or if the jdp values are
(systematically) too small.  We note that whereas for the normal 2D
image analysis, the background level has essentially no effect on the
spectral fit since the measuring aperture is so small, this is not the
case for the CC mode, for which the background of each pixel comes
from an entire column in chip coordinates.  While we cannot be
certain that the spectrum of HST-1 did not change markedly during the
months surrounding the CC observation, this seems unlikely.  Although
the CC experiment failed to provide higher accuracy for the spectral
parameters, the range of values in Table~\ref{tab:cc} is still within
the ranges for HST-1 prior to the major intensity increase.

\subsection{Fluxes and flux densities}

While most fluxes quoted from Chandra data are based on the usual
spectral analyses (sec.~\ref{sec:spectral}), we favor fluxes from
fluxmaps as being a quantity which is closer to an 'observable' and
less susceptable to model assumptions.  As described above, measuring
on flux maps achieves this objective when pileup is not a problem.  To
generate approximate fluxes from the observed keV/s measurements, we
convert keV to ergs and divide by the effective area.

Since all indications are that most of the flux falls within the 0.2
to 6 keV band, we take a nominal effective area of 400 cm$^2$ as a
first guess for the spectrum of HST-1 weighting the Chandra effective
area.  We then calculate a flux:

\smallskip
$F(0.2-6~keV) = 1.602\times10^{-9} \frac{K}{400~\times~a} 
= 4.00\times10^{-12}~\frac{K}{a}$~erg~cm$^{-2}~$s$^{-1}$
\smallskip

\noindent where K is the measured keV/s value and $a$ is a correction factor for
the (unknown) effective area which is a function of time because of
the ACIS contamination buildup.

We find the values of $a$ by comparing F(a=1) with the flux measured
on the flux maps for those observations made in 2002 and 2003 before
pileup became a serious problem.  We find a(2000.5828)=1.356, thereafter
declining to values around 1.2 by mid 2003.  Making a rough
extrapolation of a(t) permits us to make a first order correction for
the contamination problem.  We estimate that resultant uncertainties
are of order 15\%, based on the scatter of a(t) from a smooth curve.
Fluxes are listed in Table~\ref{tab:obs}.  In the same table we give
flux densities at 2 keV calculated from the flux values assuming
$\alpha$=1.5.

\section{Synchrotron Parameters\label{sec:syncpar}}

In this section we examine the physical parameters associated with the
synchrotron emission model in the same context as used in Paper I: we
use the rise time to estimate the physical diameter and the decay time
to constrain the beaming factor, assuming E$^2$ losses dominate.

For a given interval, $\Delta$t, we measure a fractional increase of
y=I$_2$/I$_1$.  Then the doubling time, dt= [(1/(y
-1)]$\times\Delta$t, and the doubling time in the jet frame is
dt$^{\prime}$ = $\delta\times$dt yr.  Thus the diameter of the
emitting region will be d $\leq$ dt$^{\prime}$/3.26 pc and the radius,
r = d/(2$\times$77) arcsec.  From inspection of fig.~\ref{fig:hst1lc}, we
select a number of intervals with large rate of increase and in
Table~\ref{tab:dt} we give the calculated doubling times.  For a
selection of values for the beaming factor, in Table~\ref{tab:size} we
give the corresponding upper limit to the sizes of the emitting region
for a characteristic doubling time of 0.14 yr.

\begin{table}[t]
\begin{center}
\caption{Doubling Times for HST-1\label{tab:dt}}
\begin{tabular}{lccc}
\tableline\tableline
Epoch Interval &  $\Delta$t & (y-1)$^{\rm a}$ &dt$^{\rm b}$\\
&  && (yrs)\\
\tableline
E-F   &     0.125   &   0.373   &   0.34 \\
I-J   &     0.090   &   0.352   &   0.26 \\
M-N   &     0.099   &   0.413   &   0.24 \\
P-Q   &     0.123   &   0.998   &   0.12 \\
X-Ya  &     0.183   &   0.413   &   0.44 \\
\tableline
\end{tabular}
\end{center}
Note. -- With a statistical error of a few percent on each
intensity measurement, we estimate the uncertainties of the doubling
times are dominated by systematic errors of order 15\%.\\
$^{\rm a}$ y is the intensity ratio, I$_2$/I$_1$.\\
$^{\rm b}$ dt is the doubling time in our frame.
\end{table}

\begin{table}[b]
\begin{center}
\caption{Size of Emitting Region for HST-1\label{tab:size}}
\begin{tabular}{rlllllll}
\tableline\tableline
$\delta$${^a}$= &  1 & 3 &5 & 9 & 12& 20 & 40 \\
\tableline
dt$^{\prime}$\tablenotemark{b}(yrs) = &   0.14 &    0.42  &   0.70  &   1.3  &   1.7  &   2.8   &   5.6  \\
diameter(pc) =    &      0.043  &    0.13     &   0.22    &   0.39 &   0.52    &   0.86    &   1.72 \\
radius(arcsec) =&    0.00028  &    0.00084  &   0.0014  &   0.0025  &   0.0034  &   0.0056  &   0.011\\
\tableline
\end{tabular}
\end{center}
Note. -- For an observed doubling time of $\approx$0.14 years,
we calculate maximum values for the source size.\\
\tablenotetext{a}{$\delta$ is the beaming factor.}
\tablenotetext{b}{dt$^{\prime}$ is the doubling time in the jet frame.}
\end{table}

We construct a spectrum for 2005Jan1 with data from
the VLA; at 220nm from the Hubble Space Telescope; and our Chandra
value for Jan 2005 from Table~\ref{tab:obs}.  This is shown in
fig.~\ref{fig:sed}.  We then convert the frequencies and flux
densities to the jet frame and solve for the usual synchrotron
parameters \citep{andre}: see Table~\ref{tab:sync}.

\begin{center}
\epsscale{1.0}
\figurenum{4}
\plotone{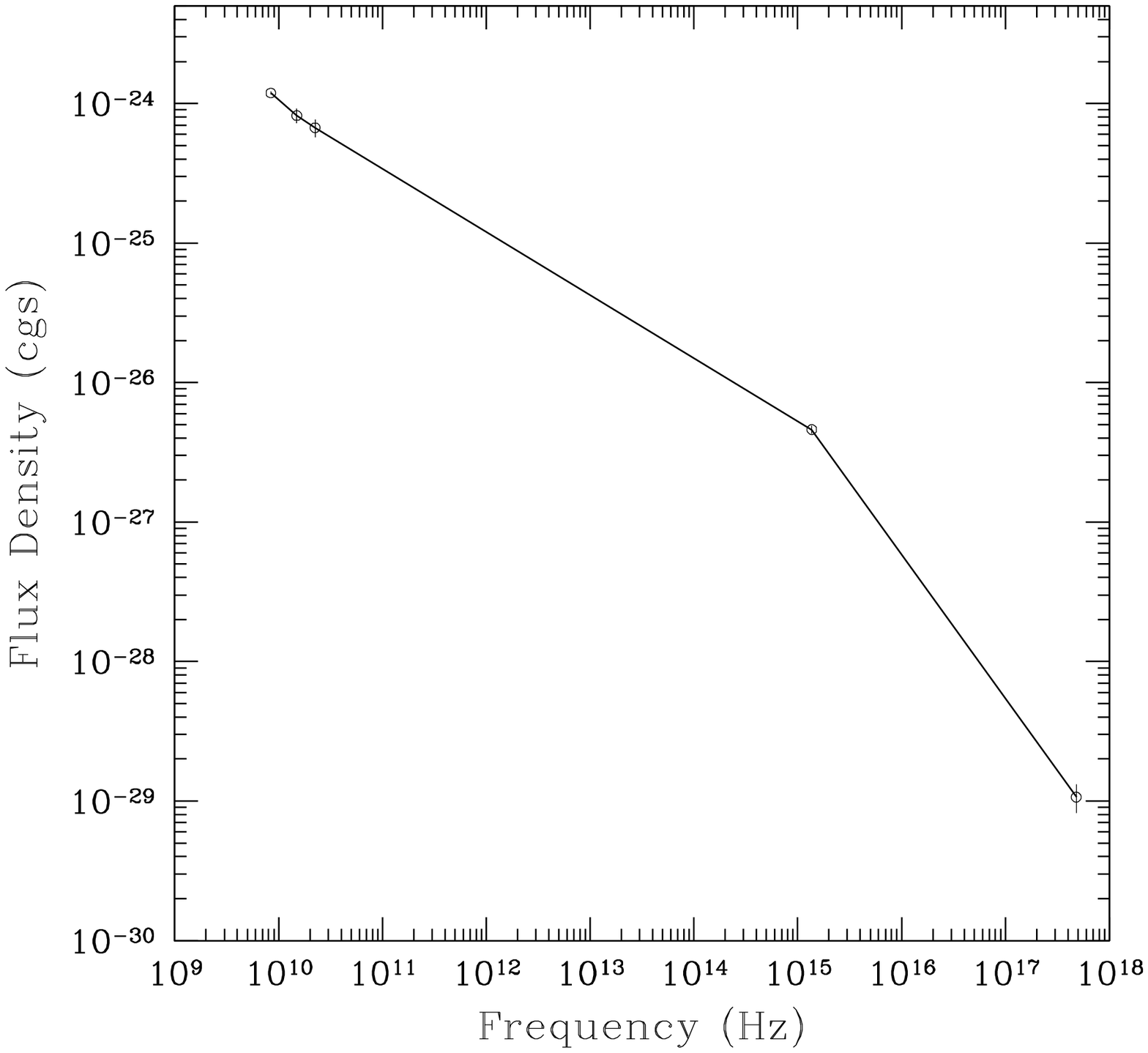}
\figcaption{The spectrum of HST-1 around 1 Jan 2005.  The radio flux
densities are from the VLA; the UV from the HST; and the Chandra
interpolated from Table~\ref{tab:obs}.  The two point spectral indices
are $\alpha_{\rm r-UV}$=0.45 (22GHz to 220nm) and $\alpha_{\rm ox}$=1.03
(220nm to 2 keV).\label{fig:sed}}
\vspace{0.1in}
\end{center}

 
\begin{table}
\begin{center}
\caption{Synchrotron Parameters for HST-1\label{tab:sync}}
\begin{tabular}{cllllcccl}
\tableline\tableline
$\delta$ & radius\tablenotemark{a} & B$_{\rm eq}$\tablenotemark{b} & 
R\tablenotemark{c} & log E$_{\rm tot}$\tablenotemark{d} & 
P\tablenotemark{e} & $\tau'_{\rm UV}$\tablenotemark{f} & 
$\tau_x^{\prime}$/d(l.y.)\tablenotemark{g} & $\tau_x$\tablenotemark{h}\\ 
 & (cm) & (gauss) &  & (ergs) & (erg s$^{-1}$) & (yrs) & (ly) & (yrs)\\
\tableline
1  &    8.3$\times10^{16}$  &  0.10   & 1.67    &  48.72   & 1.2$\times10^{42}$   & 0.015  & 0.02 &   0.0015  \\
3  &    2.5$\times10^{17}$  &  0.01   & 0.084   &  48.11   & 1.0$\times10^{41}$   &  0.91   & 0.32 &   0.030  \\
5  &    4.1$\times10^{17}$  &  0.0037 & 0.021   &  47.82   & 3.1$\times10^{40}$   &  6.03   & 1.34 &   0.10   \\
9  &    7.5$\times10^{17}$  &  0.0010 & 0.0043  &  47.50   & 8.4$\times10^{39}$   &  53.8   & 6.48 &   0.60   \\
20 &    1.6$\times10^{18}$  &  0.0002 & 0.0006  &  47.05   & 1.3$\times10^{39}$   & 894     & 50.2 &   4.47    \\
\tableline
\end{tabular}
\end{center}
Note. -- The HST-1 spectrum used is that of fig.~\ref{fig:sed}.\\
\tablenotetext{a}{The radius comes from the doubling time (Table~\ref{tab:dt}).}
\tablenotetext{b}{The standard equipartition magnetic field strength.}
\tablenotetext{c}{R is the ratio of energy densities in synchrotron
photons to that of the magnetic field.  Hence also the ratio of
synchrotron self-Compton losses to synchrotron losses.}
\tablenotetext{d}{The total energy in particles and fields required to
explain the observed synchrotron emission.}
\tablenotetext{e}{The power required to produce E$_{\rm tot}$ in time dt$^{\prime}$ (Table~\ref{tab:size}).}
\tablenotetext{f}{The halflife in the jet frame for electrons
responsible for the UV emission.  $\tau_x^{\prime}$ is a factor of
ten smaller, and these jet frame halflives can be compared to the
diameter in lightyears to see if geometrical effects or loss times dominate.}
\tablenotetext{g}{$\tau_x^{\prime}$/d(l.y.) is the ratio of the X-ray
half-life to the source diameter measured in light years.}
\tablenotetext{h}{The halflife in the observer's frame for electrons
responsible for the X-rays.}
\end{table}

We have identified the most rapid decay rates, and these are given in
Table~\ref{tab:decay}.  Estimates of the time for the intensity to
drop by a factor of two range from 0.2 to 0.3 years.  Matching 0.2 yrs
to the results of Table~\ref{tab:sync}, we find $\delta$=6.3, a value
somewhat higher than that found in Paper I and also larger than the
models reported in Table 3 of \citet{bir99} for the 6c proper motions
in HST-1.  However, the X-ray decay time is remarkably close to the
rise time (which we used to estimate the source size), so it seems
likely that both rise and fall reflect the time required to cross
the source, and not the E$^2$ half-life.  Although this is not proven,
for the remainder of this paper we adopt the view that the X-ray decay
time reflects the light travel time across HST-1.  Thus solutions for
smaller $\delta$ (e.g. 3 to 5) are favored.  Given the assumptions of
the model, $\tau_x^{\prime}$=diameter(l.y.) for $\delta\approx4.6$ (see
the next to last column in Table~\ref{tab:sync}).  For smaller
$\delta$ the lightcurve decay should be dominated by light travel
time; for larger $\delta$ the weaker magnetic field leads to longer
E$^2$ halflives.  Large $\delta$ (e.g. 20) values are disfavored
because of their large $\tau$ values and the much smaller angles to
the line of sight which would be required.

In Paper I we argued that the total emission of HST-1 did not impose
any significant drain on the total estimated energy flow down the jet.
This continues to be the case so long as $\delta$ is $\geq$~3.  The
power needed to generate the required total energy in the time dt$^{\prime}$
(the jet frame doubling time) is listed in Table~\ref{tab:sync}.
These values may be compared to the estimate of \citet{you02} for the
jet's kinetic energy flow of 3$\times~10^{42}$ erg~s$^{-1}$, based on
energetics of the interaction between the jet and the ISM which are
not based on particular jet models.  The power required to explain the
observed flare in M87 would be comparable to the jet's kinetic energy
only if there were no beaming, an untenable hypothesis.

\begin{table}
\begin{center}
\caption{Decay times for HST-1\label{tab:decay}}
\begin{tabular}{lccc}
\tableline\tableline
Epoch Interval &  $\Delta$t & y\tablenotemark{a} &
halftime\tablenotemark{b}\\
 &  (years) &   & (years) \\
\tableline
B-C   &    0.075  &  0.801     &        0.19\tablenotemark{c} \\
G-H   &    0.111  &  0.817     &        0.30\\
L-M   &    0.126  &  0.840     &        0.40\\
Ya-Yg &    0.115  &  0.840     &        0.36\\
Yh-Yi &    0.126  &  0.692     &        0.21\\
\tableline
\end{tabular}
\end{center}
Note. -- All values are in the observer's frame.
\tablenotetext{a}{y is the fractional drop of intensity: I$_2$/I$_1$}
\tablenotetext{b}{halftime=$\frac{0.5}{(1-y)}\times\Delta$t}
\tablenotetext{c}{This value was used in Paper I.}
\end{table}

\section{Implications from rise and fall timescales\label{sec:timescales}}

There is a long history supporting the notion that brightness
enhancements occur at the locations of internal shocks which produce
power law spectra of relativistic electrons via Fermi acceleration
\citep[e.g.,][]{roger1}.  Polarization data support this notion with
increased polarizations and electric vectors perpendicular to the knot
leading edge.  These are characteristic of shocks and are associated
with the flux maximum regions \citep{per99,pap2}.

If we view the shock model as a method of generating a new or
augmented power law of relativistic electrons (and increasing the
field strength), we note that there is no difficulty with the
energetics (see above) and also note that the general behaviour of the
lightcurves in fig.~\ref{fig:hst1lc} demonstrate comparable rates of
intensity increase for radio, optical, and X-rays.  This is a
pre-requisite for the shock model which relies on the power law
extending from low values of $\gamma$ up to $\approx~10^7$.
Whether the flare owes its genesis to an increase in delivered energy
from the nucleus, or to a change in the size and strength of the
shock, is not determined.

The clear prediction resulting from a new power law distribution is
that the UV emission should decay 10 times more slowly than the
X-rays, and the radio emission should persist much longer because of
the $\gamma^{-1}$ dependence of the electron half-life suffering
E$^2$ losses.  This should be observed in the lightcurves so long as
the light travel time across the source is significantly less than
the half-life.  For the synchrotron model described in
Table~\ref{tab:sync}, the estimated size (in l.y.) of the emitting
region is comparable to the halflife of electrons responsible for the
X-ray emission, but less than the predicted half-life for UV emitting
electrons.  In Paper II, we suggested that the optical decay times
were similar to those in the X-rays, but this was based on a quite
small intensity drop.  Given the marked X-ray decline starting in
2005Apr, this issue should be clarified by optical and radio data
obtained in 2005 but not yet available to us.

\section{Conclusions\label{sec:conc}}

The three lightcurves (X-rays, UV and radio; fig.~\ref{fig:hst1lc}) are
parallel to a good accuracy. This means that the input spectrum of
relativistic particles which caused the flare did not change its
shape, only its amplitude.  The ``choppiness'' in the X-ray light
curve may be caused by synchrotron losses which would be much weaker
at lower energies.

If we match the observed decay time to the halflife of electrons
producing the X-rays, we find $\delta$ a bit larger than 5.  However,
if the decay time is governed by the source size, as suggested by the
near equivalence of rise and decay times, we should choose a
$\delta$ which satisfies the condition
$\tau_x^{\prime}~\leq$~diameter(l.y.).  This requires $\delta~\leq$~5,
and that would be consistent with our results of Paper I that
$\delta~\approx$~4.

Radio and optical data now being obtained should provide the basis for
'solving' two distinct issues:  separating light travel time from
electron halflives and distinguishing between compression/expansion
versus the injection of new particles.

If the decay segments of the lightcurves show longer decay times for
lower frequencies, it will be possible to separate the light travel
time across the source from E$^2$ loss times.  From the models already
considered, probably the optical, but certainly the radio, halflives
should be considerably longer than the light travel time, so the
corresponding decay times in these lightcurves will provide a
reasonably direct estimate of the halflife against E$^2$ losses.
Since the magnetic field energy density is greater than the photon
energy density in all models considered, we will thus obtain another
estimate of the magnetic field strength from the observed halflife.

If however, all bands rise and fall together, the
compression/expansion explanation will be supported.  Given an
emitting region, be it the result of a shock or some other process, a
rather modest compression can increase the intensity by the required
factor of 50 since all electrons will receive the same percentage
energy boost and the field strength will increase.  Thus the flare
would result from a simple compression without the introduction of
'new particles'.  If the subsequent decay of intensity were to arise
from an expansion, all bands should drop equally with no delays
between bands.  Moreover, the relative rate of expansion will be
reflected in the light curve.  If the decay timescale is comparable to
the rise time, this would indicate a rapid expansion (i.e. rapid drop
in volume emissivity) whereas a decay significantly slower than the
rise would point to a slower expansion.  In the latter case, expansion
velocities can be estimated from the lightcurve with simple
assumptions.


\acknowledgments

Some of the presented data were based on observations made with the
NASA/ESA Hubble Space Telescope, obtained at the Space Telescope
Science Institute, which is operated by the Association of
Universities for Research in Astronomy, Inc., under NASA contract NAS
5-26555.  Work at SAO was supported by NASA contract NAS8-39073 and
grants GO2-3144X, GO3-4124A, and GO4-5131X.  Work at U. Maryland was
supported by NASA through LTSA grant NAG513065 to the University of
Maryland and at UMBC, the support came from NASA LTSA contracts
NAG5-9997 and NNG05-GD63G, as well as HST grants GO-9705.01 and
GO-09474.08.  We thank O. Stohlman for writing scripts and organizing
our flux measurements during the early parts of this project.
M. Nowak and J. Davis provided useful advice on ACIS pileup and the
anonymous referee is thanked for helpful comments.

Facilities:  HST(STIS), CXO(ASIS), VLA.

\end{document}